\documentclass[10pt,letterpaper,twocolumn]{article} %% two column, final layout
\usepackage{ol2}
\usepackage[draft]{hyperref}
\usepackage{amsmath}

\begin{document}

\twocolumn[ %% activate for two-column option

\title{ Weak measurement of the Goos-H$\rm\bf\ddot{a}$nchen shift}

%% For REVTeX it is possible to automate superscript and e-mail callouts with the superscriptaddress option; see REVTeX4 documentation.

\author{G. Jayaswal,$^{1}$ G. Mistura,$^1$ and M. Merano$^{1,*}$}

\address{
$^1$Dipartimento di Fisica e Astronomia G. Galilei, Università degli studi di Padova, via Marzolo 8, 35131 Padova, Italy
\\
$^*$Corresponding author: michele.merano@unipd.it
}

\begin{abstract}It is well known from quantum mechanics that weak measurements offer a means of amplifying and detecting very small phenomena. We present here the first experimental observation of the Goos-H$\rm\ddot{a}$nchen shift via a weak measurement approach. \end{abstract}


 ] %% activate for two-column option

\noindent A bounded beam of light reflected from or transmitted through a planar interface suffers diffractive corrections to the law of reflection or to Snell's law. The most prominent of these are the Goos-H$\rm\ddot{a}$nchen (GH) \cite{Goos47} and the Imbert-Fedorov (IF) \cite{Imbert72, Fedorov55} shifts that induce spatial translations of the beam in the directions parallel and perpendicular to the plane of incidence, respectively. Angular analogous to the GH \cite{Merano09} and the IF effect have recently been observed, as well as the spin Hall effect of light (SHEL) \cite{Onoda04, Kwiat08, Bliokh06}. This last one is connected to the IF shift being a separation orthogonal to the plane of incidence of the two spin components of the reflected or transmitted beam. These effects do not occur only for simple planar interfaces but were also observed or predicted for photonics crystals, waveguides or resonators \cite{Martina06, Felbacq04, Boardman07}. They are not restricted to fully spatially coherent beams but they are observed also for beams with a partial degree of spatial coherence \cite{Merano12, Wolfgang12}. They are wavelike phenomena that occur also for matter waves or acoustic waves \cite{Haan10, Merano102}.

The measurement of these optical shifts is in general a challenging task because they are tiny phenomena. A weak measurement approach has proven to be successful for the observations of these effects. This technique is an optical analogue \cite{Sudarshan89} of the quantum weak measurement concept introduced in ref. \cite{Aharonov88}. In a remarkable experiment, Hosten and Kwiat \cite{Kwiat08} where the firsts that applied this experimental scheme to the measurement of optical beam shifts. They reported the first experimental observation of the SHEL. This observation was also confirmed by other experiments \cite{Qin09}. Theoretical analysis of the weak measurement approach for the observation of optical beam shifts were reported in ref. \cite{Gotte12}.

In this letter we present the first experimental observation of the GH effect via a weak measurement scheme. The ``weak measuring device'' \cite{Sudarshan89} is a prism which introduces a small lateral displacement $D_{p}$ ($D_{s}$) for the $p$ ($s$) polarization of a Gaussian beam in total internal reflection (TIR). We consider a Cartesian reference frame with $y$ in the vertical direction, $z$ in the propagation direction and $x$ fixed as consequence. A polarizer and an analyzer select the initial and the final linear polarizations to be at angles $\alpha$ and $\beta$ with respect to the horizontal in the $xy$ plane. The weak measurement scheme works as follow \cite{Sudarshan89, Ritchie91}: if the polarizer and the analyzer are set to $\alpha$ = $\rm \pi \over 4$ and $\beta$ = $\alpha$ +$ \pi \over 2$ + $\epsilon$ ($\epsilon$ $\ll$ 1) respectively, the emerging beam is laterally shifted of a quantity $1 \over 2$$\cdot$$\Delta_{GH}\cdot \rm{cot}(\epsilon)$, where $\Delta_{GH}$ = $D_{p}$ - $D_{s}$. We measure the separation ($\Delta_{GH}\cdot \rm{cot}(\epsilon)$) in between the two beams corresponding to the two polarizations settings $\alpha$  = $ \pi \over 4$  and $\beta$ = $\alpha$ +$ \pi \over 2$ $\pm$ $\epsilon$. The small beam displacement $\Delta_{GH}$ introduced by the GH effect is thus amplified of a factor $\rm{cot}(\epsilon)$.

We test this in our setup (Fig.1). We collimate light from a single-mode fiber-coupled 826 nm laser diode with a 20X objective. In order to reduce problems connected to multiple reflections in the prism, a beam expander is used to generate a collimated Gaussian beam with a beam waist $w_{0}$ = 260 $\mu m$. A polarizer (P1) is set at $\alpha$  = $ \pi \over 4$ . The beam then suffers TIR in a $\rm 45^\circ - 90^\circ - 45^\circ$ prism (BK7, n = 1,51 at 826 nm), The analyzer consists of a quarter wave plate ($\lambda$/4), an half wave plate ($\lambda$/2) and a polarizer (P2). A quadrant detector (QD), mounted on a translation stage, is used to measure the relative beam displacement. It can be replaced by a CCD for observing the profile of the beam emerging from the analyzer.

\begin{figure}
\includegraphics{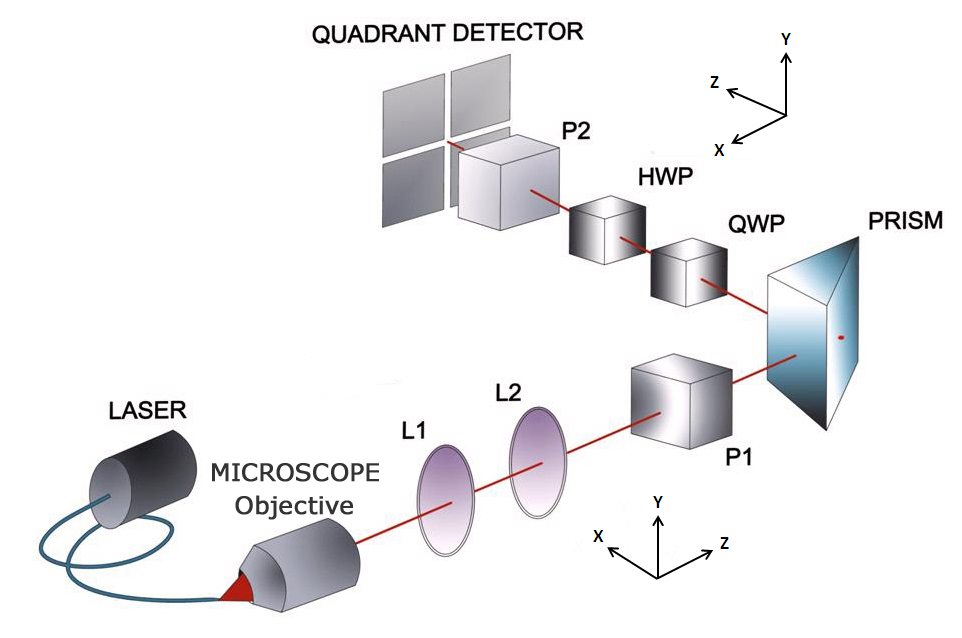}
\caption{Setup for the weak measurement of the GH shift. The polarizer P1 is followed by an
analyser composed of QWP (quarter wave plate), HWP (half wave plate) and polarizer (P2). L1 and
L2 are the lenses of the beam expander.}
\end{figure}

We first set P2 to $\beta$ = $\alpha$ +$ \pi \over 2$ and we set the $\lambda$/4 and the $\lambda$/2 in order to minimize the power transmitted to the QD. We then turn P2 first to $\beta$ = $\alpha$ +$ \pi \over 2$ + $\epsilon$ and then to $\beta$ = $\alpha$ +$ \pi \over 2$ - $\epsilon$  and we measure the beams separation S ( S = $\Delta_{GH}\cdot \rm{cot}(\epsilon)$ ) with the QD. The angle $\epsilon$ is chosen to be $10^{-2}$ rad. This value of $\epsilon$ is a convenient compromise in between transmitted power $P_{t} =10^{-4} P$ (where $P$ is the maximum power that can be transmitted through P2) and the lateral displacement amplification ($\rm{cot}(\epsilon)$ =100). With this choice, the beam profile of the transmitted beam is still Gaussian (left side of fig. 2), the only observable effect is the lateral translation. For completeness, in the same figure (right side) we repot the beam profile corresponding to crossed polarizer ($\beta$ = $\alpha$ +$ \pi \over 2$). The transmitted beam is not anymore Gaussian but has two peaks separated by approximately $2^{1/2}$w where w is the beam waist expected for the input Gaussian beam in the CCD plane \cite{Ritchie91}.  

\begin{figure}
\includegraphics{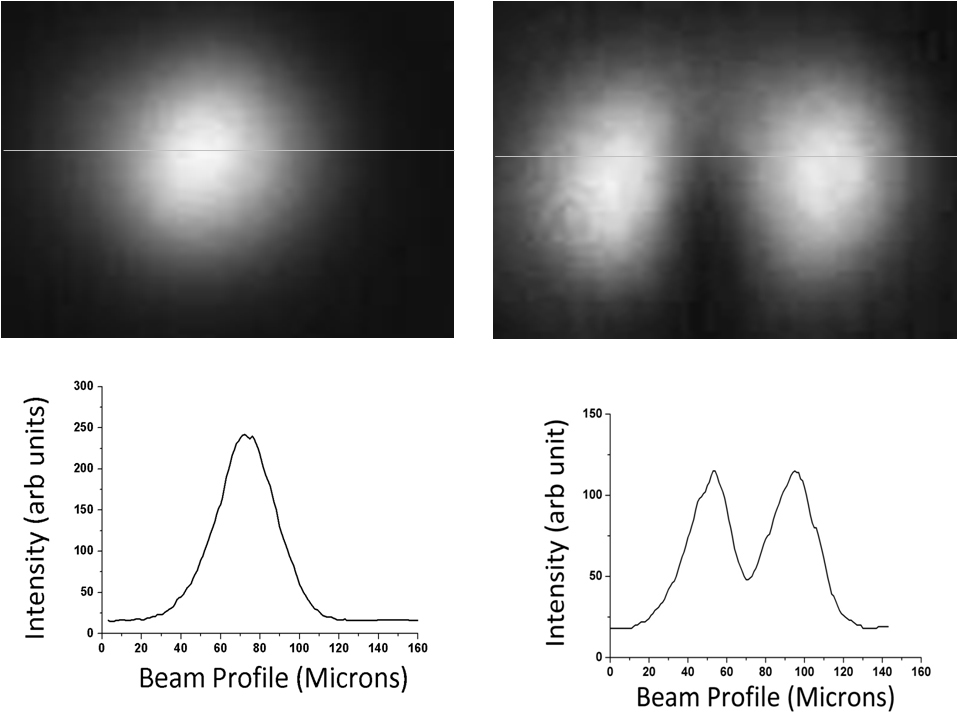}
\caption{Left: beam profile for ($\epsilon$=  $10^{-2}$ rad). Right: beam profile for perfect crossing ($\epsilon$ = 0 rad).}
\end{figure}

Figure 3 shows our experimental data. The horizontal axis is the angle of incidence, the vertical axis is the GH shift of a $p$ polarized beam with respect to a $s$ polarized beam. The line in the figure represents the theoretical predictions for a Gaussian beam derived from ref. \cite{Aiello08}. Data correspond to the GH shift $\Delta_{GH}$ derived from measurements. The small gap in the experimental data around 45$^\circ$ of incidence is due to unavoidable problems with multiple reflections in the prims.  

\begin{figure}
\includegraphics{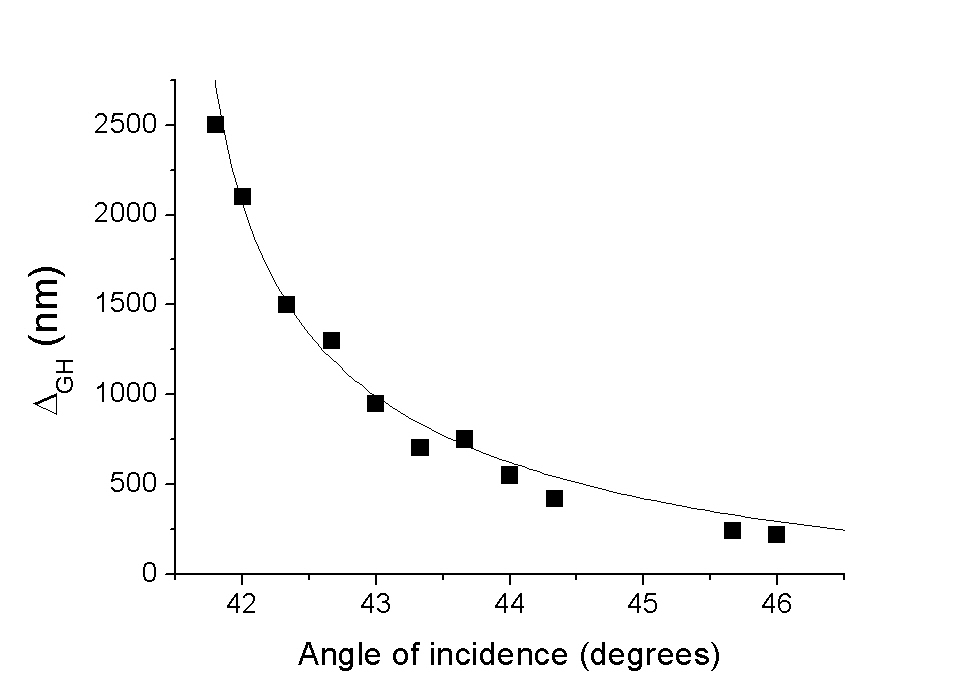}
\caption{Experimental data (squares) are compared with the theoretical predictions for the GH
shift in total internal reflection.}
\end{figure}

The agreement in between theory and experiment is very good. This is certainly a surprising result because in our theoretical analysis we used only results from ref. \cite{Ritchie91} for the weak measurements approach. In that paper the beams (the analogous of the $p$ and $s$ components in our case) emerge in phase from the ``weak measuring device''. In total internal reflection the $p$ and $s$ components undergo phase jumps $\delta_{p}$ and $\delta_{s}$ of different amounts that depend on the angle of incidence. If we follow ref. \cite{Sudarshan89} the $p$ and the $s$ components of the beam emerging from TIR will be \\ 
\begin{displaymath}
\begin{vmatrix}
\frac{1}{\sqrt{2}}e^{-i\frac{\delta}{2}}\exp{\Big[-\frac{(x-D_{p})^2}{w_{0}^{2}}\Big]} \\ \\
\frac{1}{\sqrt{2}}e^{i\frac{\delta}{2}}\exp{\Big[-\frac{(x-D_{s})^2}{w_{0}^{2}}\Big]} 
\end{vmatrix}
\end{displaymath} \\ 
where $\delta$ = $\delta_{p}$ - $\delta_{s}$ is the phase difference between $p$ and $s$. (Multiplication by a complex exponential affects only absolute phase, not relative phase or intensity, and hence not polarization form \cite{Kliger}.) We introduce the $\lambda$/4 and the $\lambda$/2 to compensate for the relative phase and we note that in this particular case the Jones matrix representing their effect is a diagonal one:\\ 
\begin{displaymath}
\begin{vmatrix}
e^{i\frac{\delta}{2}} & 0 \\
0 & e^{-i\frac{\delta}{2}}
\end{vmatrix}
\end{displaymath}
\\

A non-diagonal matrix would mix the two polarization components. The beam emerging from our retarder has the following form\\
\begin{displaymath}
\begin{vmatrix}
\frac{1}{\sqrt{2}}\exp{\Big[-\frac{(x-D_{p})^2}{w_{0}^{2}}\Big]} \\ \\
\frac{1}{\sqrt{2}}\exp{\Big[-\frac{(x-D_{s})^2}{w_{0}^{2}}\Big]} 
\end{vmatrix}
\end{displaymath} 
\\
and it is identical to formula 40 of ref. \cite{Sudarshan89} from which the formalism of \cite{Ritchie91} is derived. This is the reason why theory and experiments agree so well. 

\begin{align*}
f(x)=\cos(\beta)\cos(\alpha)\exp{\Big[-\frac{(x-D_{p})^2}{w^{2}}\Big]} \\ +\sin({\beta})\sin({\alpha})\exp{\Big[-\frac{(x-D_{s})^2}{w^{2}}\Big]}
\end{align*} 
This can be written as (ref. \cite{Sudarshan89}, formula 42)
\begin{align*}
f(x)=\cos({\alpha}+{\beta}){\phi}(x-a; {\epsilon}, w_{0}, \Delta_{GH}) 
\end{align*}
where $a=\frac{1}{2}(D_{p}+D_{s}) $, $\tan(\epsilon) \simeq \epsilon$ (in our case $\epsilon =10^{-2}$ rad), and $\phi$ is given by
\begin{align*}
\phi(x)=\frac{1}{2}\Big[(1+\epsilon)\exp{\Big[-\frac{(x-\frac{1}{2}\Delta_{GH})^2}{w^{2}}\Big]}\\ - (1-\epsilon)\exp{\Big[-\frac{(x+\frac{1}{2}\Delta_{GH})^2}{w^{2}}\Big]}\Big]
\end{align*} 
this last form makes it evident that that the wave function is a superposition of two Gaussians, centered on $x=\pm\frac{1}{2}\Delta_{GH}$. It is now simple to show that for $\frac{1}{2}\Delta_{GH}/w\ll \epsilon \ll 1$ the two Gaussians interfere desstructively to produce a single Gaussian, the centroid of which is shifted by the weak value $\frac{1}{2}\Delta_{GH}/ \epsilon \simeq \frac{1}{2}\Delta_{GH}\cot(\epsilon)$ \cite{Ritchie91}. 

We find our result particularly nice because by choosing a fixed value for $\epsilon =\pm 10^{-2}$ rad, we obtain a neat amplification of the GH curve predicted by the standard theory of the GH effect \cite{Aiello08}. We simply multiply by a factor 100 the GH shift due to TIR for any angle of incidence. This is not always the case for weak measurements applied to beam shifts. For instance in ref \cite{Kwiat08, Qin09} the opposite displacements of the two spin components actually depend on the input polarization state that, for experimental reasons, is varied when the angle of incidence change. 

The theoretical support for our experimental results is based on refs \cite{Sudarshan89, Ritchie91, Aiello08}. Important experimental and theoretical papers considering weak measurement of the beam shifts have been reported recently \cite{Kwiat08, Qin09, Qin11, Gorodetski12, Dennis12, Gotte12}. Anyway most of them \cite{Kwiat08, Qin09, Gorodetski12} deal with SHEL and not with the GH effect. The nice work of G$\rm\ddot{o}$tte and Dennis \cite{Dennis12, Gotte12} treats together the case for the GH, IF and the SHEL. The work of Qin et al. \cite{Qin11} treats the angular GH shift and the SHEL. Unfortunately these theories do not apply to our experiment because they do not explicitly consider an analyser made of $\lambda$/4, $\lambda$/2 and a polarizer.

In summary, we have demonstrated that the weak measurement is a valuable approach for observing the GH shift in TIR. It allows for a neat amplication of the effect. Although our results are strictly valid only for TIR we expect that they can be extended experimentally and theoretically  (for instance  to the case of metallic mirrors \cite{Merano07}). This technique, based on an amplification scheme, can be particularly interesting in applications where the GH shift is applied for sensing \cite{Hesselink06}.

\bibliography{letter}

\begin{thebibliography}{10}
\newcommand{\enquote}[1]{``#1''}

\bibitem{Goos47}
F.~Goos and H.~$\rm H\ddot{\textbf{a}}nchen$, \enquote{{Ein neuer und
  fundamentaler Versuch zur Totalreflexion},} Ann.\ Phys. \textbf{1}, 333
  (1947).

\bibitem{Imbert72}
C.~Imbert, \enquote{{Calculation and Experimental Proof of the Transverse Shift
  Induced by Total Internal Reflection of a Circularly Polarized Light Beam},}
  Phys. Rev. D \textbf{5}, 787 (1972).

\bibitem{Fedorov55}
F.~I. Fedorov, \enquote{{K Teorii Polnogo Otrazheniya},} Dokl. Akad. Nauk. Nauk
  SSSR \textbf{105}, 465 (1955).

\bibitem{Merano09}
M.~Merano, A.~Aiello, M.~P. van Exter, and J.~P. Woerdman, \enquote{{Observing
  angular deviations in the specular reflection of a light beam},} Nat. Photon.
  \textbf{3}, 337 (2009).

\bibitem{Onoda04}
M.~Onoda, S.~Murakami, and N.~Nagaosa, \enquote{{Hall effect of light},} {Phys.
  Rev. Lett.} \textbf{{93}}, 083901 ({2004}).

\bibitem{Kwiat08}
O.~Hosten and P.~Kwiat, \enquote{{Observation of the Spin Hall Effect of Light
  via Weak Measurements},} Science \textbf{319}, 787 (2008).

\bibitem{Bliokh06}
K.~Y. Bliokh and Y.~P. Bliokh, \enquote{{Conservation of Angular Momentum,
  Transverse Shift, and Spin Hall Effect in Reflection and Refraction of an
  Electromagnetic Wave Packet},} Phys. Rev. Lett. \textbf{96}, 073903 (2006).

\bibitem{Martina06}
H.~Schomerus and M.~Hentschel, \enquote{{Correcting ray optics at curved
  dielectric microresonator interfaces: Phase-space unification of Fresnel
  filtering and the Goos-Hanchen shift},} {Phys. Rev. Lett.} \textbf{{96}},
  {243903} ({2006}).

\bibitem{Felbacq04}
D.~Felbacq and R.~Sma$\hat{a}$li, \enquote{{Bloch modes dressed by evanescent
  waves and the generalized Goos-Hanchen effect in photonic crystals},} {Phys.
  Rev. Lett.} \textbf{{92}}, {193902} ({2004}).

\bibitem{Boardman07}
K.~L. Tsakmakidis, A.~D. Boardman, and O.~Hess, \enquote{{`Trapped rainbow'
  storage of light in metamaterials},} {Nature} \textbf{{450}}, {397} ({2007}).

\bibitem{Merano12}
M.~Merano, G.~Umbriaco, and G.~Mistura, \enquote{{Observation of nonspecular
  effects for Gaussian Schell-model light beams},} {Phys. Rev. A}
  \textbf{{86}}, {033842} ({2012}).

\bibitem{Wolfgang12}
W.~Loffler, A.~Aiello, and J.~P. Woerdman, \enquote{{Spatial Coherence and
  Optical Beam Shifts},} {Phys. Rev. Lett.} \textbf{{109}}, {213901} ({2012}).

\bibitem{Haan10}
V.-O. de~Haan, J.~Plomp, T.~M. Rekveldt, W.~H. Kraan, A.~A. van Well, R.~M.
  Dalgliesh, and S.~Langridge, \enquote{{Observation of the Goos-Hanchen Shift
  with Neutrons},} {Phys. Rev. Lett.} \textbf{{104}}, {010401} ({2010}).

\bibitem{Merano102}
M.~Merano, N.~Hermosa, A.~Aiello, and J.~P. Woerdman, \enquote{{Demonstration
  of a quasi-scalar angular Goos-Hanchen effect},} {Opt. Lett} \textbf{{35}},
  {3562--3564} ({2010}).

\bibitem{Sudarshan89}
I.~Duck, P.~Stevenson, and E.~Sudarshan, \enquote{{THE SENSE IN WHICH A WEAK
  MEASUREMENT OF A SPIN-1/2 PARTICLES SPIN COMPONENT YIELDS A VALUE 100},}
  {Phys. Rev. D} \textbf{{40}}, {2112--2117} ({1989}).

\bibitem{Aharonov88}
Y.~Aharonov, D.~Albert, and L.~Vaidman, \enquote{{HOW THE RESULT OF A
  MEASUREMENT OF A COMPONENT OF THE SPIN OF A SPIN-1/2 PARTICLE CAN TURN OUT TO
  BE 100},} {Phys. Rev. Lett.} \textbf{{60}}, {1351--1354} ({1988}).

\bibitem{Qin09}
Y.~Qin, Y.~Li, H.~He, and Q.~Gong, \enquote{{Measurement of spin Hall effect of
  reflected light},} {Opt. Lett.} \textbf{{34}}, {2551--2553} ({2009}).

\bibitem{Gotte12}
J.~B. $\rm G\ddot{o}tte$ and M.~R. Dennis, \enquote{{Generalized shifts and
  weak values for polarization components of reflected light beams},} {New J.
  Phys.} \textbf{{14}} ({2012}).

\bibitem{Ritchie91}
N.~Ritchie, J.~Story, and R.~Hulet, \enquote{{REALIZATION OF A MEASUREMENT OF A
  WEAK VALUE},} {Phys. Rev. Lett.} \textbf{{66}}, {1107--1110} ({1991}).

\bibitem{Aiello08}
A.~Aiello and J.~P. Woerdman, \enquote{{Role of beam propagation in
  Goos-Hanchen and Imbert-Fedorov shifts},} {Opt. Lett.} \textbf{{33}}, {1437}
  ({2008}).

\bibitem{Kliger}
D.~S. Kliger, J.~W. Lewis, and C.~E. Randall, \emph{Polarized light in optics
  and spectroscopy} (Academic Press, 1990), 1st ed.

\bibitem{Qin11}
Y.~Qin, Y.~Li, X.~Feng, Y.-F. Xiao, H.~Yang, and Q.~Gong, \enquote{{Observation
  of the in-plane spin separation of light},} {Opt. Exp.} \textbf{{19}},
  {9636--9645} ({2011}).

\bibitem{Gorodetski12}
Y.~Gorodetski, K.~Y. Bliokh, B.~Stein, C.~Genet, N.~Shitrit, V.~Kleiner,
  E.~Hasman, and T.~W. Ebbesen, \enquote{{Weak Measurements of Light Chirality
  with a Plasmonic Slit},} Phys. Rev. Lett. \textbf{109}, 013901 (2012).

\bibitem{Dennis12}
M.~R. Dennis and J.~B. Goette, \enquote{{The analogy between optical beam
  shifts and quantum weak measurements},} {New J. Phys.} \textbf{{14}}
  ({2012}).

\bibitem{Merano07}
M.~Merano, A.~Aiello, G.~W. `t~Hooft, M.~P. van Exter, E.~R. Eliel, and J.~P.
  Woerdman, \enquote{{Observation of Goos-Hanchen shifts in metallic
  reflection},} {Opt. Exp.} \textbf{{15}}, {15928--15934} ({2007}).

\bibitem{Hesselink06}
X.~Yin and L.~Hesselink, \enquote{{Goos-H$\rm \ddot{\textbf{a}}$nchen shift
  surface plasmon resonance sensor},} Appl.\ Phys.\ Lett. \textbf{89}, 261108
  (2006).

\end{thebibliography}
\bibliographystyle{osajnl}

\end{document}